\documentclass[aps,showpacs,manuscript,12pt]{revtex4}
\usepackage{amssymb}
\usepackage{amsmath}
\usepackage{graphicx}

\begin{document}
\title{\textbf{Axisymmetric gravitational MHD equilibria in the presence of plasma
rotation$^{\S }$}}
\author{C. Cremaschini$^{a}$, A. Beklemishev$^{b}$, J. Miller$^{c,d}$ and M.
 Tessarotto$^{e,f}$}
\affiliation{\ $^{a}$Department of Astronomy, University of
Trieste, Italy, $^{b}$ Budker Institute of Nuclear Physics,
Novosibirsk, Russia, $^{c}$International School for Advanced
Studies, SISSA and INFN, Trieste,  Italy,  $^{d}$Department of
Physics (Astrophysics), University of Oxford, Oxford, U.K.,
$^{e}$Department of Mathematics and Informatics, University of
Trieste, Italy, $^{f}$Consortium of Magneto-fluid-dynamics,
University of Trieste, Italy}
\begin{abstract}

In this paper, extending the investigation developed in an earlier
paper (Cremaschini et al., 2008), we pose the problem of the
kinetic description of gravitational Hall-MHD equilibria which may
arise in accretion disks (AD) plasmas close to compact objects.
When intense EM and gravitational fields, generated by the central
object, are present, a convenient approach can be achieved in the
context of the Vlasov-Maxwell description. In this paper the
investigation is focused primarily on the following two aspects:

1) the formulation of the kinetic treatment of G-Hall-MHD
equilibria. Based on the identification of the relevant first
integrals of motion, we show that an explicit representation can
be given for the equilibrium kinetic distribution function. For
each species this is represented as a superposition of suitable
generalized Maxwellian distributions;

2) the determination of the constraints to be placed on the fluid
fields for the existence of the kinetic equilibria. In particular,
this permits a unique determination of the functional form of the
species number densities and of the fluid partial pressures, in
terms of suitably prescribed flux functions.
\end{abstract}
\date{\today }
\maketitle

%$^{\star }$} \url{http://cmfd.univ.trieste.it}.

%%%%%%%%%%%%%%%%%%%%%%%%%%%%%%%%%%%%%%%%%%%%
%% MAINMATTER
%%%%%%%%%%%%%%%%%%%%%%%%%%%%%%%%%%%%%%%%%%%%

\section{Introduction}

An open issue in astrophysics is understanding the dynamics of
accretion disk (AD) plasmas occurring near compact objects (black
holes, neutron
stars, etc.) \cite{Chandrasekhar,Morozov,Catto1999,Tasso,McClements,Ghanbari}%
. It is well known that this problem must be generally formulated
in the context of kinetic theory for the proper description of
macroscopic plasma dynamics. In fact, unless direct information
about the fluid fields which characterize the plasma is available
(for example from theory, particle numerical simulations or direct
observations), a purely fluid treatment of these plasmas may pose
serious difficulties. These are related to the notorious
impossibility of uniquely defining consistent closure conditions
for the fluid equations. \ A convenient solution can, however, be
reached by adopting a kinetic approach. An important motivation
for this investigation is, in fact, that of avoiding the closure
problem characteristic of fluid equations (see the related
discussion in Paper I, Cremaschini et al., 2008). A first step in
this direction, realized in this paper, is related to the
investigation of kinetic equilibria (in this context, what is
meant by an "equilibrium" is stationary-flow solution). The
purpose of this paper is to investigate, in particular, kinetic
axi-symmetric gravitational Hall-MHD (G-Hall-MHD) equilibria\
occurring in AD plasmas arising around compact stars, which are
locally characterized by the presence of a family of nested
axi-symmetric toroidal magnetic surfaces $\left\{ \psi (\mathbf{r}%
)=const\right\} ,$ $\psi $ denoting the so-called poloidal flux.
These plasmas are expected to be collisionless and characterized
by the presence of intense EM fields, as well as - at the same
time - by a strong gravitational field generated by the central
object. Extending the investigation developed in an earlier paper
(see Paper I) and based on a fluid description, in this paper we
intend to analyze, in particular, the kinetic constraints placed
on the relevant fluid fields by the kinetic treatment of
G-Hall-MHD equilibria, i.e., the requirement that the kinetic
distribution function is itself a stationary solution of the
relevant kinetic equation. Ignoring possible weakly-dissipative
and time-dependent effects as well as assuming the possible
presence of a radial flow velocity in the AD, we shall assume - in
particular - that the kinetic distribution function and the EM
fields associated with the plasma obey the system of
Vlasov-Maxwell equations. In addition, the effect of the
gravitational field produced by the central object is treated by
the introduction of a pseudo-Newtonian gravitational potential
which permits to retain weakly relativistic effects on the AD
plasma dynamics. In such a case, in general, the form of the
equilibrium kinetic distribution remains completely arbitrary. The
only restriction on its form, besides its (strict) positivity and
the assumption of its suitable smoothness in the relevant
phase-space, is obviously due to the requirement that it must be a
function only of the independent first integrals of motion. As
such, it is always possible to represent it as superpositions of
suitable generalized Maxwellian distributions (GMD). In this paper
we wish to analyze the constraints placed on them by the
assumption of the existence of kinetic equilibria of this type. We
intend to prove, in particular, that:

\begin{enumerate}
\item The assumption of kinetic equilibrium determines uniquely the
functional form of the number densities, flow velocities and
temperatures carried by each GMD, which are found to depend on
appropriate flux functions to be suitably prescribed.

\item In particular, we intend to prove that the flow velocities carried by
each GMD are species-dependent but coincide, for each GMD
belonging to the same species.

\item In addition, the related angular velocities [$\Omega _{s}$] are
constant on each toroidal magnetic surface $\left\{ \psi (\mathbf{r}%
)=const\right\}$, while the form of $\Omega _{s}$ is uniquely
prescribed in terms of normal derivatives of suitable flux-surface
averages of the electrostatic and gravitational pseudo-Newtonian
potentials.
\end{enumerate}

\section{Kinetic G-Hall-MHD equilibria}

As a starting point, we require that AD plasma admits fluid
equilibrium, i.e., a G-Hall-MHD equilibrium in the sense defined
in Paper I. \ In particular, we assume that the equilibrium
magnetic field $\mathbf{B}$
admits, at least locally, a family of magnetic surfaces $\left\{ \psi (%
\mathbf{r})\right\} \equiv \left\{ \psi
(\mathbf{r})=const\right\},$  all mutually nested and bounded,
which are represented by smooth toroidal surfaces, $\psi $
denoting the poloidal magnetic flux of $\mathbf{B}$ (see Paper I).
By assumption, these surfaces and all equilibrium fluid fields are
axi-symmetric. This means that the relevant dynamical variables
characterizing the plasma are required to be independent of the
toroidal
angle $\varphi,$ when referred to a set of cylindrical coordinates $%
(R,\varphi ,z)$. In particular, the axis $R=0$ can be identified
with the principal axis of the toroidal surfaces. The assumption
of the existence of a family of nested magnetic surfaces $\left\{
\psi (\mathbf{r})\right\} $ implies that locally a set of magnetic
coordinates ($\psi ,\varphi ,\vartheta $) can be defined, with
$\vartheta $ denoting, in particular, a
curvilinear angle-like coordinate on a magnetic surface $\psi (\mathbf{r}%
)=const.$ For definiteness we shall assume, furthermore, that the
plasma is: a) \emph{non-relativistic}, in the sense that - for an
inertial observer at rest with respect to the central object and
in the considered region of the AD - the species flow velocities
are much smaller, in magnitude, than the speed of light in vacuum;
\ b) \emph{collisionless}, so that the mean free path of the
particles of the plasma is assumed to be much larger than the
largest characteristic scale length of the plasma; c) \ the plasma
can be treated in the \emph{pseudo-Newtonian approximation},
whereby the gravitational force produced by the central object is
treated by means of an appropriate pseudo-Newtonian gravitational
potential acting on the plasma.

Let us now introduce the key assumption that the G-Hall-MHD
equilibrium also corresponds to a kinetic equilibrium, namely that
the kinetic distribution functions characterizing the plasma
species are stationary solutions of the Vlasov kinetic equation.
We stress that the condition of kinetic equilibrium is here
assumed to apply perhaps only in a local sense, i.e., in an
appropriate sub-domain of the AD where the plasma can be treated,
in particular, according to the previous assumptions a)-c). In
particular, for greater generality and in contrast with respect to
Paper I, here we shall allow that the equilibrium \ magnetic field
admits a non-vanishing toroidal component ($\mathbf{B}_{T}$),
possibly produced by currents (of the AD plasma) located far from
the considered region. This means that the kinetic distribution
functions must be an exact first integral of motion. For
definiteness we shall consider here a plasma formed by at least
two species of charged particles: one species of ions and one of
electrons. Regarding the specific form of the species-equilibrium
kinetic distribution functions $f_{k}$ (where $k=i,e$ in the case
of a two-species plasma formed by one species of ions and the
electrons), we shall assume that they can be approximated by a
superposition of suitable GMD's $f_{\ast s},$
namely for $k=i,e$%
\begin{equation}
f_{k}=\sum\limits_{s\in I_{k}}f_{\ast s}
\end{equation}%
where $I_{i}$ and $I_{e}$ are suitable sets of indices. Thus, in
this sense for the kinetic description of the plasma, several
sub-species $s=1,..,n$ can in principle be distinguished, each one
by assumption defined in such a way that:

I) the kinetic distribution $f_{\ast s}$ is a first integral of motion$%
, $ in the sense that:%
\begin{equation}
\frac{d}{dt}\ln f_{\ast s}=0.  \label{adiabatic invariant}
\end{equation}

II) for each $s=1,n$ the equilibrium kinetic distribution
function, denoted by $f_{\ast s},$ is - in a suitable asymptotic
sense to be defined later - a drifted local Maxwellian
distribution of the form:
\begin{equation}
f_{Ms}=\frac{n_{s}}{\pi ^{3/2}v_{ths}^{3}}\exp \left\{
-\frac{M_{s}\left( \mathbf{v}-\mathbf{V}_{s}\right)
^{2}}{2T_{s}}\right\} .  \label{Maxwellian}
\end{equation}

This means that there exists an infinitesimal dimensionless parameter $%
\varepsilon $ such that
\begin{equation}
f_{\ast s}=f_{Ms}\left[ 1+o(\varepsilon )\right] .
\label{approximation-0}
\end{equation}

Here the notation is standard, thus $n_{s},$ $T_{s}$ and
$\mathbf{V}_{s}$ are respectively the sub-species number density,
temperature and flow velocity. Moreover, $v_{ths}=\left\{
2T_{s}/M_{s}\right\} ^{1/2}$ is the
thermal velocity and $\mathbf{b}$ is the unit vector of the magnetic field $%
\mathbf{B}(\mathbf{r,}t)$. \

III) each distribution function $f_{Ms}$ (for $s=1,n$) is an
approximate stationary solution of the Vlasov equation, in the
sense that, consistent
with the requirement (\ref{approximation-0}), it must give%
\begin{equation}
\frac{d}{dt}\ln f_{Ms}=0+o(\varepsilon ).
\end{equation}

It can be proved, as a consequence, that the following
\emph{kinetic constraints} must be imposed on the fluid fields:

\begin{enumerate}
\item $\mathbf{V}_{s}(\mathbf{r})$ can be be identified with a toroidal flow
velocity%
\begin{equation}
\mathbf{V}_{s}=\mathbf{e}_{\varphi }R\Omega _{s}\mathbf{,}
\label{toridal velocity}
\end{equation}%
where the toroidal angular frequency takes the form $\Omega
_{s}=\Omega
_{s}(\psi ),$ i.e., it is constant on magnetic surfaces $\psi (\mathbf{r}%
)=const;$

\item the number density $n_{s}$ is:
\begin{equation}
n_{s}=n_{os}\exp \left\{ -\frac{\widetilde{S}_{s}}{T_{s}}\right\}
, \label{2}
\end{equation}%
where
\begin{equation}
S_{s}=Z_{s}e\phi +U_{Grav}-\frac{M_{s}}{2}V_{s}^{2}  \label{2b}
\end{equation}%
and $\widetilde{A}=A-\left\langle A\right\rangle .$ Here the
notation is standard (see Paper I). In particular, $\phi $ and
$U_{Grav}$ are respectively the equilibrium electrostatic
potential and an effective Pseudo-Newtonian gravitational
potential. Moreover, $\left\langle
A\right\rangle $ is the $\psi -$surface average (defined on a flux surface $%
\psi (\mathbf{r})=const.$) of function $A(\mathbf{r}),$ by
assumption
independent of the toroidal angle $\varphi ,$ which is defined as $%
\left\langle A\right\rangle =\xi ^{-1}\oint d\vartheta A(\mathbf{r}%
)/\left\vert \mathbf{B}\cdot \nabla \vartheta \right\vert ,$ with
$\xi $ denoting $\xi \equiv \oint d\vartheta /\left\vert
\mathbf{B}\cdot \nabla \vartheta \right\vert ;$

\item ignoring possible slow spatial dependencies, the pseudo-density $%
n_{os} $ and the temperature $T_{s}$ are flux functions, namely
they depend
only on the poloidal flux $\psi $ so that:%
\begin{equation}
n_{os}=n_{os}(\psi ),  \label{3}
\end{equation}%
\begin{equation}
T_{s}=T_{s}(\psi );  \label{5}
\end{equation}

\item the toroidal angular velocity $\Omega _{s}(\psi )$ can always be
identified with the flux function
\begin{equation}
\Omega _{s}(\psi )=\Omega (\psi )+\Delta \Omega _{s}(\psi ),
\label{toroidal angular frequency}
\end{equation}%
where\ $\Omega (\psi )=c\frac{\partial }{\partial \psi
}\left\langle \Phi \right\rangle $ and $\Delta \Omega _{s}(\psi
)=\frac{1}{Z_{s}e}\left\langle U_{Grav}\right\rangle $ are the
toroidal angular velocities driven, respectively, by the
electrostatic and effective gravitational potentials.
\end{enumerate}

Let us first determine the equilibrium distribution function
$f_{\ast s}.$ For this purpose, we first notice that it must be
necessarily a function only of the independent first integrals of
motion and/or of the relevant
independent adiabatic invariants (at least for the leading orders in $%
\varepsilon $). In axisymmetry these are well known \cite%
{Catto1987,Tessarotto1992} in particular the canonical momentum $%
p_{\varphi s}$ and the total particle energy, $E_{s}=\frac{M_{s}\mathbf{v}%
^{2}}{2}+\frac{Z_{s}e}{\varepsilon }\Phi +\frac{1}{\varepsilon
}U_{grav}.$ In the following we assume that the plasma is strongly
magnetized \ in the
sense that for all species $s=e,i$ %
\begin{equation}
\frac{r_{Ls}}{L}\sim \varepsilon ,  \label{epsilon}
\end{equation}%
where $r_{Ls}=v_{\perp ths}/\Omega _{cs}$ is the species average
Larmor radius and $\Omega _{cs}=Z_{s}eB/M_{s}c$ is the species
Larmor frequency. Here we shall assume that the plasma is subject
to intense electrostatic and
gravitational fields (defined respectively by the electrostatic potential $%
\Phi $ and the effective gravitational potential $U_{grav}$), thus requiring%
\begin{equation}
\frac{T_{s}}{Z_{s}e\Phi }\sim \frac{T_{s}}{U_{grav}}\sim
o(\varepsilon ). \label{ordering}
\end{equation}%
As a consequence, the magnetic flux $\psi ,$ as well as $\Phi $
and the
effective gravitational potential $U_{grav}$ are all considered to be of order $%
1/\varepsilon ,$ while the species thermal velocities, \ i.e., $%
v_{\parallel ths}$ and $v_{\perp ths}$ (for $s=e,i$), are assumed to be of order $%
o(\varepsilon ^{0}).$ Constructing the adiabatic invariants in the
following we shall consider in particular "thermal" test-particles for which $%
\left\vert \mathbf{v}\right\vert $ is of the same order as
$v_{ths}$ (for $s=1,n$)$.$ It follows that, thanks to axisymmetry,
the canonical momentum conjugate to the ignorable (toroidal) angle
$\varphi $ is by assumption a first integral. Taking into account
the previous asymptotic orderings, it is given by
\begin{equation}
p_{\varphi s}=M_{s}R\mathbf{v\cdot }\widehat{e}_{\varphi }+\frac{Z_{s}e}{%
\varepsilon c}\psi \equiv \frac{Z_{s}e}{\varepsilon c}\psi _{\ast
s}
\end{equation}%
Similarly at equilibrium the test particle total energy, $E_{s}=\frac{M_{s}%
\mathbf{v}^{2}}{2}+\frac{Z_{s}e}{\varepsilon }\Phi +\frac{1}{\varepsilon }%
U_{grav}$ is by assumption a first integral of motion. It follows
that
the dynamical variable%
\begin{equation*}
H_{\ast s}=E_{s}-\frac{Z_{s}e}{\varepsilon c}\psi _{\ast s}\Omega
_{s}(\psi _{\ast s})
\end{equation*}%
is manifestly a first integral of motion too. We notice that
introducing the velocity vector $\mathbf{V}_{\ast
s}=\mathbf{e}_{\varphi }R\Omega _{s}(\psi
_{\ast s})$, it follows that $H_{\ast s}$ is also given by $H_{\ast s}=E_{Rs}(%
\mathbf{V}_{s})-\frac{Z_{s}e}{\varepsilon c}\psi \Omega _{s}(\psi _{\ast s})-%
\frac{M_{s}V_{\ast s}^{2}}{2},$ where%
\begin{equation}
E_{Rs}(\mathbf{V}_{\ast s})=\frac{M_{s}(\mathbf{v}-\mathbf{V}_{\ast s}%
\mathbf{)}^{2}}{2}+\frac{Z_{s}e}{\varepsilon }\Phi +\frac{1}{\varepsilon }%
U_{grav},
\end{equation}%
so that $E_{Rs}(\mathbf{V}_{s})$ can be interpreted as the total
energy relative to the reference frame moving with velocity
$\mathbf{V}_{\ast s}.$
Next, we note that invoking the orderings (\ref{epsilon}),(\ref{ordering}%
) and assuming $\left\vert \mathbf{v}\right\vert /v_{ths}\sim
o(\varepsilon ^{0})$, the variables $\psi _{\ast s}$ and $H_{\ast
s}$ can be conveniently
approximated as%
\begin{equation}
\psi _{\ast s}\cong \psi \left[ 1+o(\varepsilon )\right]
\label{app-1}
\end{equation}%
\begin{equation}
H_{\ast s}=H_{s}\left[ 1+o(\varepsilon ^{2})\right]  \label{app-2}
\end{equation}%
where
\begin{equation}
H_{s}=E_{Rs}(\mathbf{V}_{s})-\frac{M_{s}V_{s}^{2}}{2}-\frac{Z_{s}e}{c}\frac{%
\psi }{\varepsilon }\Omega _{s}(\psi ),
\end{equation}%
and $E_{Rs}(\mathbf{V}_{s})$ is the total energy relative to the
reference frame moving with the (fluid) velocity
$\mathbf{V}_{s}=\mathbf{e}_{\varphi }R\Omega _{s}(\psi )$ [see
Eq.(\ref{toridal velocity})].

Thus, a possible definition for the equilibrium distributions
$f_{\ast s}$ is provided by the distribution function
\begin{equation}
f_{\ast s}=\frac{\widehat{n}_{\ast s}}{\pi ^{3/2}\left( 2T_{\ast
s}/M_{s}\right) ^{3/2}}\exp \left\{ -\frac{H_{\ast s}}{T_{\ast
s}}\right\}
\end{equation}%
(\textit{generalized Maxwellian distribution}), where
$\widehat{n}_{\ast s}$ and $T_{\ast s}$ are defined as
$\widehat{n}_{\ast s}=\widehat{n}_{s}(\psi _{\ast s})$ and
$T_{\ast s}=T_{s}(\psi _{\ast s}).$  It clearly follows that:

\begin{itemize}
\item for all $s=1,n,$ $f_{\ast s}$ is a function of first integrals and
hence, by construction, is itself a first integral of motion. This
result holds, in principle, for an arbitrary function $\Omega
_{s}(\psi _{\ast s})$ (assumed to be suitably smooth);

\item the kinetic constraints (\ref{toridal velocity})-(\ref{toroidal
angular frequency}) are necessarily identically satisfied, by
suitable definition of the functions $\widehat{n}_{s}(\psi _{\ast
s})$ and $\Omega _{s}(\psi _{\ast s})$;

\item invoking the asymptotic orderings (\ref{epsilon}),(\ref{ordering}) and
$\left\vert \mathbf{v}\right\vert /v_{ths}\sim o(\varepsilon ^{0}),$ Eq.(\ref%
{app-2}) is satisfied for $f_{Ms}$ defined by
Eq.(\ref{Maxwellian}).
\end{itemize}

To prove these statements, we invoke Eqs. (\ref{app-1}) and
(\ref{app-2}) so
that $f_{\ast s}$ can be written%
\begin{equation}
f_{\ast s}=\frac{\widehat{n}_{s}(\psi )}{\pi ^{3/2}\left(
2T_{s}(\psi )/M_{s}\right) ^{3/2}}\exp \left\{
-\frac{H_{s}}{T_{s}(\psi )}\right\} \left[ 1+o(\varepsilon
)\right] .  \label{first approx}
\end{equation}%
By making use of the expressions%
\begin{equation}
n_{os}(\psi )=\widehat{n}_{s}(\psi )\exp \left\{-
\frac{\left\langle S\right\rangle _{s}}{T_{s}}\right\} ,
\label{pos-1}
\end{equation}%
it follows that Eq.(\ref{first approx}) leads to
(\ref{approximation-0}),
while Eq.(\ref{pos-1}) implies (\ref{2}). \ Finally, we notice that Eq.(\ref%
{toroidal angular frequency}) \ can also be obtained in an
equivalent way from the momentum balance equation for the $s-$th
species [see Eqs.(1) and (2) in Paper I)]. In the present case,
this gives
\begin{equation}
\frac{q_{s}n_{s}}{\varepsilon }\left( -\nabla \phi \mathbf{+E}%
_{s}^{(d)}\right) -\frac{q_{i}}{\varepsilon
c}n_{s}\mathbf{V}_{s}\times \mathbf{B=0,}  \label{momentum-i}
\end{equation}%
where $\mathbf{E}_{s}^{(d)}$ is a suitable diamagnetic electric
field (see
Eq.(3), Paper I). \ It follows $\mathbf{V}_{s}\equiv \Omega R\mathbf{e}%
_{\varphi }=V_{\parallel }\mathbf{b+}\frac{c}{B}\left( -\nabla \phi \mathbf{%
+E}_{s}^{(d)}\right) \times \mathbf{b}$. As a consequence, taking
the scalar products term by term with $\mathbf{b,}$ $\mathbf{e}%
_{\varphi }$ and $\nabla \psi $ it follows that
\begin{eqnarray}
&&\left. V_{\parallel }=\Omega R\frac{B_{T}}{B},\right.  \\
&&\left. \Omega R\frac{B_{P}^{2}}{B^{2}}=\nabla \psi \cdot \left(
\nabla
\phi -\mathbf{E}_{s}^{(d)}\right) \right.  \\
&&\left. \left( -\nabla \phi \mathbf{+E}_{s}^{(d)}\right) \mathbf{\cdot B}%
_{p}=0\right. ,
\end{eqnarray}%
which \ proves Eq.(\ref{toroidal angular frequency}). This result
shows that in such a case, i.e., for \ kinetic G-Hall-MHD
equilibria which hold under the validity of the previous
asymptotic orderings, no bifurcation arises (in contrast to the
general case of fluid equilibria, considered in Paper I). This is
due to the functional form of the number density provided by
(\ref{2}). \ As a consequence, as indicated above, in the present
case the fluid velocity ($\mathbf{V}_{s}$) is uniquely determined
and is identical for each sub-species belonging, respectively, to
the ion or electron species. In addition, the presence of a finite
(equilibrium) toroidal magnetic field is in principle permitted.
Although its possible origin has not been explicitly discussed
here, we can always assume that it is produced by "external"
currents (i.e., located outside the domain of local existence of
the kinetic equilibrium). Finally, another interesting consequence
concerns the form of the generalized Grad-Shafranov equation,
which in
the present case becomes%
\begin{equation}
\frac{1}{R}\frac{\partial ^{2}\psi }{\partial z^{2}}+\frac{1}{R}\frac{%
\partial ^{2}\psi }{\partial R^{2}}=-\frac{4\pi R}{c}\left[ \rho \Omega
+\sum\limits_{s}Z_{s}en_{s}\Delta \Omega _{s}\right] .
\end{equation}%
This shows that in kinetic G-Hall-MHD equilibria, the electric
current density, producing the self-generated poloidal magnetic
field, contains two terms: a) the first one driven by the $\psi
$-derivative of the average electrostatic potential is
proportional to the local charge density of the plasma $\rho $
(Hall effect); b) the second one due to the gravitational
potential, occurs even in the case of a quasi-neutral plasma
($\rho \cong 0$).

\section{Conclusions}

In this paper, kinetic G-Hall-MHD equilibria have been
investigated. In particular, the kinetic equilibrium has been
determined imposing three main assumptions:

1) that a fluid G-Hall-MHD equilibrium exists which is
characterized locally
by a family of nested magnetic surfaces $\left\{ \psi (\mathbf{r}%
)=const\right\} $ represented by axi-symmetric nested tori;

2) requiring that the species equilibrium kinetic distribution
function can be represented by a superposition of suitable local
equilibrium distributions $f_{\ast s}$ (for $s=1,n$);

3) imposing that the equilibrium distribution function can be
expressed in terms of first integrals of the motion.

Regarding the second assumption we remark that: a) experimental
observations of collisionless astrophysical plasmas (for example,
the solar wind) are consistent with this type of representation;
b) it is always possible to represent a kinetic distribution
function by an appropriate superposition of drifted local
Maxwellian distributions (here replaced by the equilibrium
distributions $f_{\ast s})$. Nevertheless, as shown in this paper,
the assumption of kinetic equilibrium invoked here has been
obtained by imposing the validity of suitable \textit{kinetic
constraints }on these distributions. As in Paper I, no assumption
of local quasi-neutrality has been invoked, thus permitting the
consistent treatment of Hall effects. In addition, allowance has
been made for the inclusion of relativistic effects, taken into
account by means of an effective gravitational potential
($U_{grav}$), and the presence of a finite toroidal magnetic
field. Finally, we remark that the third assumption actually rules
out the treatment of possible more general kinetic equilibria,
which may be expressed in terms of adiabatic invariants to be
established based on the
so-called gyrokinetic theory of particle dynamics in strong EM fields \cite%
{Cremaschini2008c}. The main results of the present theory
include: 1) the treatment of kinetic equilibria in the presence of
strong EM and gravitational fields, i.e., in particular within the
validity of the asymptotic orderings\
(\ref{epsilon}),(\ref{ordering});\ 2) the evaluation of the
species-dependent toroidal angular velocity ($\Omega _{i}$), which
- to leading order in the asymptotic parameter $\varepsilon ,$ is
proved to depend on the gradient of the $\psi -$average
electrostatic and effective gravitational potentials; 3) the proof
that for this type of kinetic
equilibrium, no bifurcation occurs in the ion toroidal angular velocity ($%
\Omega _{i}$). The theory developed here applies to plasmas
subject to the action of an intense magnetic field, partly
self-generated by the plasma itself. The theory permits also the
treatment of intense electric and gravitational fields, including
the possible action of relativistic effects produced by the
presence of a massive central object. In particular, the
equilibrium distribution $f_{\ast s}$ has been determined by
adopting a gyrokinetic approach. This permits $f_{\ast s}$ to be
constructed in such a way that it is - at the same time -a first
integral of the motion and is also close, in an appropriate
asymptotic sense, to a drifted local Maxwellian distribution.
Based on the present theory, G-Hall-MHD equilibria can be
investigated, in principle, utilizing either standard numerical
solution methods or by a perturbative solution method based on a
power series expansion near to the equatorial plane ($z=0$).
Detailed results will be presented elsewhere
\cite{Cremaschini2008}. The results presented in this paper appear
to be potentially significant both for the numerical evaluation of
AD plasma equilibria and for the physical interpretation of the
theory. We believe, in fact, that applications of the present
theory can be useful for the theoretical investigation of AD
phenomenology and for gaining a better understanding of the
related basic physical processes.\

\section*{Acknowledgments}
This work has been developed in cooperation with the CMFD Team,
Consortium for Magneto-fluid-dynamics (Trieste University,
Trieste, Italy), within the framework of the MIUR (Italian
Ministry of University and Research) PRIN Programme: {\it Modelli
della teoria cinetica matematica nello studio dei sistemi
complessi nelle scienze applicate}. Support is acknowledged (by
A.B.) from ICTP (International Center for Theoretical Physics,
Trieste, Italy) and the University of Trieste, Italy, (by M.T.)
from COST Action P17 (EPM, {\it Electromagnetic Processing of
Materials}) and GNFM (National Group of Mathematical Physics) of
INDAM (Italian National Institute for Advanced Mathematics).

\section*{Notice}
$^{\S }$ contributed paper at RGD26 (Kyoto, Japan, July 2008).

\end{document}